\newcommand{\Pder}[2]{\frac{\partial #1}{\partial #2}}
\newcommand{\Psecder}[3]{\frac{\partial^2 #1}{\partial #2 \partial #3}}

\def\Tr{\mathop{\rm Tr}}
\newcommand{\prob}[1]{\mathbb{P}\left[#1\right]}
\newcommand{\expec}[2]{\mathbb{E}_{#1}\left[#2\right]}
\newcommand{\cov}[2]{\mathrm{Cov}\left(#1,#2\right)}
\newcommand{\rd}{{\rm d}}
\renewcommand{\vec}[1]{\boldsymbol #1}

\newtheorem{lemma}{Lemma}
\newtheorem{theorem}{Theorem}

\documentclass[10pt]{paper}
\usepackage{a4wide}
\usepackage{amsfonts}
\usepackage{amsmath}
\usepackage{amssymb}
\usepackage{url}
\usepackage{epsfig}
\usepackage{graphics}
\usepackage{graphicx}

\begin{document}

\setlength{\parindent}{0mm}

\title{Asymptotic fingerprinting capacity\\ for non-binary alphabets}
\author{Dion Boesten and Boris \v{S}kori\'{c}\\
Eindhoven University of Technology}

\date{ }

\maketitle

\begin{abstract}
\noindent{\it
We compute the channel capacity of non-binary 
fingerprinting under the Marking Assumption, in the limit of large coalition size~$c$.
The solution for the binary case was found by Huang and Moulin.
They showed that asymptotically, the capacity is $1/(c^2 2\ln 2)$, the interleaving attack is optimal
and the arcsine distribution is the optimal bias distribution.

In this paper we prove that the asymptotic capacity for general alphabet size $q$
is $(q-1)/(c^2 2\ln q)$.
Our proof technique does not reveal the optimal attack or bias distribution.
The fact that the capacity is an increasing function of $q$ shows that
there is a real gain in going to non-binary alphabets.}
\end{abstract}
\section{Introduction}

\subsection{Collusion resistant watermarking}
Watermarking provides a means for tracing the origin and
distribution of digital data. Before distribution of digital content,
the content is modified by applying an imperceptible watermark (WM), 
embedded using a watermarking algorithm. 
Once an unauthorized copy
of the content is found, it is possible to trace those users who
participated in its creation. 
This process is known as `forensic watermarking'.
Reliable tracing requires resilience against attacks that aim
to remove the WM.  Collusion attacks,
where several users cooperate, are a
particular threat: differences between their versions of the content
tell them where the WM is located.
Coding theory has produced a number of
collusion-resistant codes. The resulting system has two
layers: 
The coding layer determines which message to embed and protects against collusion attacks. 
The underlying watermarking layer hides symbols of the code in segments\footnote{
The `segments' are defined in a very broad sense. They may be coefficients in any
representation of the content (codec).
}
of the content.
The interface between the layers is usually specified in terms of the \textit{Marking
Assumption}, which states that the colluders are able to
perform modifications only in those segments where they received different WMs.  
These segments are called detectable positions.

Many collusion resistant codes have been proposed in the literature. 
Most notable is the Tardos code \cite{Tardos},
which achieves the asymptotically optimal proportionality $m\propto c^2$,
with $m$ the code length. 
Tardos introduced a two-step stochastic procedure for generating binary codewords:
(i) For each segment a bias is randomly drawn from some distribution $F$.
(ii) For each user independently, a 0 or 1 is randomly drawn for each segment
using the bias for that segment.
This construction was generalized to larger alphabets in \cite{symmetric}.

\subsection{Related work: channel capacity}

In the original Tardos scheme \cite{Tardos}
and many later improvements and generalisations 
(e.g. \cite{SVCT,symmetric,BlayerTassa,Nuida,Nuida_DCC2009,FuronEM,CombinedDigit,XFF}), 
users are found to be innocent or guilty via an `accusation sum',
a sum of weighted per-segment contributions, computed for each user separately.
The discussion of achievable performance was 
greatly helped by the onset of an information-theoretic treatment
of anti-collusion codes.
The whole class of bias-based codes can be treated as a maximin game
between the watermarker and the colluders \cite{ABD,Moulin2008,HM_ISIT2009},
independently played for each segment,
where the payoff function is the mutual information
between the symbols $x_1,\ldots,x_c$ handed to the colluders and the
symbol $y$ produced by them.
In each segment (i.e. for each bias)
the colluders try to minimize the payoff function using 
an attack strategy that depends on the (frequencies of the) received symbols $x_1,\ldots,x_c$. 
The watermarker tries to maximize the
average payoff over the segments by setting the bias distribution~$F$.

It was conjectured \cite{HM_ISIT2009} that the binary capacity is asymptotically given by
$1/(c^2 2\ln 2)$.
The conjecture was proved in \cite{AmiriTardos,HuangMoulin}.
Amiri and Tardos \cite{AmiriTardos} developed an accusation scheme 
(for the binary case)
where candidate coalitions get a score related 
to the mutual information between their symbols and~$y$.
This scheme achieves capacity but is computationally very expensive.
Huang and Moulin \cite{HuangMoulin} proved for the large-$c$ limit
(in the binary case)
that the interleaving attack and Tardos's arcsine distribution are optimal.

\subsection{Contributions and outline}

We prove for alphabet size $q$
that the asymptotic fingerprinting capacity is $\frac{q-1}{c^2 2\ln q}$.
Our proof makes use of the fact that the value of the maximin game can be found
by considering the minimax game instead (i.e. in the reverse order).
This proof does not reveal the asymptotically optimal collusion strategy and
bias distribution of the maximin game.

In Section~\ref{sec:preliminaries} we introduce notation, 
discuss the information-theoretic payoff game
and present lemmas that will be used later.
In Section~\ref{sec:analysis} we analyze the properties of the payoff function
in the large-$c$ limit.
We solve the minimax game in Section~\ref{sec:solution}.
In Section~\ref{sec:discussion} we discuss the benefits of larger alphabets.

\section{Preliminaries}
\label{sec:preliminaries}

\subsection{Notation}
We use capital letters to represent random variables, and lowercase letters to their realizations. Vectors are denoted in boldface and the components of a vector $\vec x$ are written as $x_i$.
The expectation over a random variable $X$ is denoted as $\mathbb{E}_X$.
The mutual information between $X$ and $Y$ is denoted by $I(X ; Y)$, and the mutual information conditioned on a third variable $Z$ by $I(X;Y | Z)$.
The base-$q$ logarithm is written as $\log_q$ and the natural logarithm as $\ln$. If $\vec p$ and $\vec \sigma$ are two vectors of length $n$ then by $\vec{p}^{\vec \sigma}$ we denote $\prod_{i=1}^n p_i^{\sigma_i}$. If $c$ is a positive integer and $\vec \sigma$ is a vector of length $n$ of nonnegative integers with sum equal to $c$ then ${c \choose \vec \sigma}$ denotes the multinomial coefficient $\frac{c!}{\sigma_1! \sigma_2! \dots \sigma_n!}$. The standard Euclidean norm of a vector $\vec x$ is denoted by $\| \vec x \|$. The Kronecker delta of two variables $\alpha$ and $\beta$ is denoted by $\delta_{\alpha\beta}$. A sum over all possible outcomes of a random variable $X$ is denoted by $\sum_x$.
In order not to
clutter up the notation we will often omit the set to which $x$ belongs when it is clear from the context.

\subsection{Fingerprinting with per-segment symbol biases}

Tardos \cite{Tardos} introduced the first fingerprinting scheme that achieves optimality
in the sense of having the asymptotic behavior $m\propto c^2$.
He introduced a two-step stochastic procedure for generating the codeword matrix~$X$.
Here we show the generalization to non-binary alphabets \cite{symmetric}.
A Tardos code of length $m$ for a number of users $n$ over the alphabet $\mathcal{Q}$ of size $q$
is a set of $n$ length-$m$ sequences of symbols from $\mathcal{Q}$
arranged in an $n \times m$ matrix $X$. The codeword for a user $i \in \{1,\ldots,n\}$ is the $i$-th row in $X$.
The symbols in each column $j \in \{1,\ldots,m\}$ are generated in the following way.
First an auxiliary bias vector $\vec P^{(j)} \in [0,1]^q$ with $\sum\limits_{\alpha} P^{(j)}_{\alpha} = 1$ is generated independently for each column $j$, from a distribution $F$.
(The $\vec P^{(j)}$ are sometimes referred to as `time sharing' variables.)
The result $\vec p^{(j)}$ is used to generate each entry $X_{ij}$ of column $j$ independently:
$\prob{X_{ij} = \alpha} = p^{(j)}_{\alpha}$.
The code generation has independence of all columns and rows.

\subsection{The collusion attack}
Let the random variable $\Sigma_\alpha^{(j)}\in\{0,1,\ldots,c\}$ denote the number of colluders
who receive the symbol $\alpha$ in segment $j$. It holds that $\sum_\alpha\sigma_\alpha^{(j)}=c$ for all $j$.
From now on we will drop the segment index $j$, since all segments are independent.
For given $\vec p$, the vector $\vec\Sigma$ is multinomial-distributed,
\begin{equation}
	\Lambda_{\vec\sigma|\vec p}\triangleq {\rm Prob}[\vec\Sigma=\vec\sigma|\vec P=\vec p]
	={c\choose\vec\sigma}\vec p^{\vec\sigma}.
\end{equation}
The colluders' goal is to produce a symbol $Y$ that does not incriminate them.
It has been shown that it is sufficient to consider a probabilistic per-segment (column)
attack which does not distinguish between the different colluders.
Such an attack then only depends on $\vec\Sigma$, and
the strategy can be completely described by a set of probabilities
$\theta_{y \mid \vec\sigma} \in [0,1]$, which are defined as:
\begin{align}
	\theta_{y \mid \vec\sigma} & \triangleq {\rm Prob}[Y = y \mid \vec \Sigma = \vec \sigma].
\label{def:piratestrategy}
\end{align}
For all $\vec\sigma$, conservation of probability gives $\sum_y \theta_{y \mid \vec \sigma} = 1 $.
Due to the Marking Assumption, $\sigma_\alpha=0$ implies $\theta_{\alpha|\vec\sigma}=0$
and $\sigma_\alpha=c$ implies $\theta_{\alpha|\vec\sigma}=1$.
The so called {\em interleaving attack} is defined as $\theta_{\alpha|\vec\sigma}=\sigma_\alpha/c$.

\subsection{Collusion channel and fingerprinting capacity}
The attack can be interpreted as a noisy channel with input $\vec \Sigma$ and output $Y$.
A capacity for this channel can then be defined,
which gives an upper bound on the achievable code rate of a reliable fingerprinting scheme.
The first step of the code generation, drawing the biases $\vec p$,
is not considered to be a part of the channel.
The fingerprinting capacity $C_c(q)$ 
for a coalition of size $c$ and alphabet size $q$ is equal 
to the optimal value of the following two-player game:
\begin{equation}
	C_c(q) = \max\limits_F \min\limits_{\vec \theta}
	\frac{1}{c} I(Y; \vec \Sigma \mid \vec P)
	= \max\limits_F \min\limits_{\vec \theta} \frac{1}{c}
	\int F(\vec p) I(Y; \vec \Sigma \mid \vec P = \vec p) d^q \vec p. \label{eq:fingerprintinggame}
\end{equation}
Here the information is measured in $q$-ary symbols.
Our aim is to compute the fingerprinting capacity $C_c(q)$ in the limit 
($n\to\infty$, $c \to \infty$).

\subsection{Alternative mutual information game}
\label{sec:alternative}
The payoff function of the game~(\ref{eq:fingerprintinggame}) is the mutual information
$I(Y; \vec \Sigma \mid \vec P)$.
It is convex in $\vec \theta$ (see e.g. \cite{CoverThomas}) and linear in $F$.
This allows us to apply Sion's minimax theorem (Lemma~\ref{lem:sionsminimax}), yielding
\begin{align}
	\max\limits_F \min\limits_{\vec \theta} I(Y; \vec \Sigma \mid \vec P)
	&= \min\limits_{\vec \theta} \max\limits_F I(Y; \vec \Sigma \mid \vec P)
\label{eq:alternativefingerprintinggame1}
	\\
	&= \min\limits_{\vec \theta} \max\limits_{\vec p} I(Y; \vec \Sigma \mid \vec P = \vec p) \label{eq:alternativefingerprintinggame2}
\end{align}
where the last equality follows from the fact that the maximization over $F$ in~(\ref{eq:alternativefingerprintinggame1}) results in a delta distribution located at the maximum of the payoff function. The game~(\ref{eq:fingerprintinggame}) is what happens in reality, but by solving the alternative game (\ref{eq:alternativefingerprintinggame2})
we will obtain the asymptotic fingerprinting capacity.

\subsection{Useful Lemmas}
The following lemmas will prove useful for our analysis of the asymptotic fingerprinting game.

\begin{lemma}[Sion's minimax theorem \cite{Sion}]
\label{lem:sionsminimax}
Let $\cal X$ be a compact convex subset of a linear topological space and 
$\cal Y$ a convex subset of a linear topological space.
Let $f:\cal X\times \cal Y\to\mathbb R$ be a function with

\begin{itemize}
\item $f(x,\cdot)$ upper semicontinuous and quasiconcave on $\cal Y$, $\forall x\in \cal X$
\item $f(\cdot,y)$ lower semicontinuous and quasi-convex on $\cal X$, $\forall y\in \cal Y$
\end{itemize}
then $\min_{x\in\cal X}\max_{y\in\cal Y} f(x,y)=\max_{y\in\cal Y}\min_{x\in\cal X}f(x,y)$.
\end{lemma}

\begin{lemma}\label{lem:symmetricmatrix}
Let $M$ be a real $n \times n$ matrix.
Then $M^T M$ is a symmetric matrix
with nonnegative eigenvalues. Being symmetric, $M^T M$ has mutually orthogonal eigenvectors.
Furthermore, for any two eigenvectors $\vec v_1 \perp \vec v_2$ of $M^T M$  we have $M \vec v_1 \perp M \vec v_2$.
\end{lemma}

{\it Proof:}
$M^T M$ is symmetric because we have $(M^T M)^T=M^T (M^T)^T=M^T M$.
For an eigenvector $\vec v$ of $M^T M$, corresponding to eigenvalue $\lambda$, 
the expression $\vec v^T M^T M \vec v$ can on the one hand
be evaluated to $\vec v^T \lambda \vec v=\lambda\|\vec v\|^2$, and on the other hand
to $\|M\vec v\|^2\geq 0$.
This proves that $\lambda\geq 0$.
Finally, any symmetric matrix has an orthogonal eigensystem.
For two different eigenvectors $\vec v_1$, $\vec v_2$ of $M^T M$, with $\vec v_1\perp\vec v_2$,
the expression $\vec v_1^T M^T M \vec v_2$ can on the one hand be evaluated to
$\vec v_1^T \lambda_2 \vec v_2=0$, and on the other hand to $(M\vec v_1)^T(M\vec v_2)$.
This proves $M \vec v_1 \perp M \vec v_2$.
\hfill$\square$

\begin{lemma}
\label{lem:continuousmapping}
Let $\cal V$ be a set that is homeomorphic to a (higher-dimenional) ball.
Let $\partial{\cal V}$ be the boundary of $\cal V$.
Let $f: {\cal V} \to {\cal V}$ be a differentiable function such that 
$\partial\cal V$ is surjectively mapped to $\partial\cal V$. 
Then $f$ is surjective.
\end{lemma}
{\it Proof sketch:}
A differentiable function that surjectively maps the edge $\partial\cal V$ to itself
can deform existing holes in $\cal V$ but cannot create new holes.
Since $\cal V$ does not contain any holes, neither does $f(\cal V)$.
\hfill$\square$

\begin{lemma}[Arithmetic Mean - Geometric Mean (AM-GM) inequality]
\label{lem:amgminequality}
For any $n \in \mathbb{N}$ and any list $x_1,x_2,\ldots,x_n$ of nonnegative real numbers it holds that
$\frac1n\sum_{i=1}^n x_i \geq \sqrt[n]{x_1 x_2 \dots x_n}$.
\end{lemma}

\section{Analysis of the asymptotic fingerprinting game}
\label{sec:analysis}
\subsection{Continuum limit of the attack strategy}

As in \cite{HuangMoulin} we assume that the attack strategy satisfies the following condition
in the limit $c\to\infty$.
There exists a set of bounded and twice differentiable functions $g_y: [0,1]^q \to [0,1]$,
with $y \in \mathcal{Q}$, such that
\begin{enumerate}
\item
$g_\alpha(\vec\sigma/c)=\theta_{\alpha|\vec\sigma}$ for all $\alpha$, $\vec\sigma$
\item
$x_\alpha=0$ implies $g_\alpha(\vec x)=0$
\item
$\sum_\alpha x_\alpha=1$ implies $\sum_\alpha g_\alpha(\vec x)=1$.
\end{enumerate}

\subsection{Mutual information}
We introduce the notation
$\tau_{y | \vec p} \triangleq {\rm Prob}[Y\!=\!y | \vec P\!=\!\vec p]$ 
$=\sum_{\vec \sigma} \theta_{y | \vec \sigma}\Lambda_{\vec \sigma | \vec p}$
$= \expec{\vec \Sigma|\vec P=\vec p}{\theta_{y | \vec \Sigma}}$.
The mutual information can then be expressed as:
\begin{align}
	I(Y; \vec \Sigma \mid \vec P) &=
	 \sum\limits_y \sum\limits_{\vec \sigma}
	\theta_{y \mid \vec \sigma}\Lambda_{\vec \sigma \mid \vec p} 
	\log_q \left(\frac{\theta_{y \mid \vec \sigma}}{\tau_{y \mid \vec p}}\right)
\label{eq:mutualinformationasafunctionofp}
\end{align}
where we take the base-$q$ logarithm because 
we measure information in $q$-ary symbols.
Using the continuum assumption on the strategy we can write
\begin{align}
	I(Y; \vec \Sigma \mid \vec P = \vec p)
	&= \sum\limits_y \sum\limits_{\vec \sigma}
	\Lambda_{\vec \sigma \mid \vec p} g_y(\frac{\vec\sigma}c)
	\log_q \left( \frac{g_y(\vec\sigma/c)}
	{\expec{\vec \Sigma|\vec P=\vec p}{g_y(\vec \Sigma/c)}}\right).
\label{eq:mutualinformationintermsofg}
\end{align}

\subsection{Taylor approximation and the asymptotic fingerprinting game}
For large $c$, the multinomial-distributed variable $\vec \Sigma$ tends towards its mean $c \vec p$
with shrinking relative variance.
Therefore we do a Taylor expansion\footnote{
Some care must be taken in using partial derivatives $\partial/\partial p_\beta$ of $\vec g$.
The use of $\vec g$ as a continuum limit of $\vec\theta$
is introduced on the hyperplane $\sum_\alpha p_\alpha=1$, but writing down
a derivative forces us to define $\vec g(\vec p)$ outside the hyperplane as well.
We have a lot of freedom to do so, which we will exploit in Section~\ref{sec:freedominchoosinggamma}.
}
of $\vec g$ around the point $\frac{\vec \sigma}{c} = \vec p$:
\begin{equation}
	g_y\!\!\left(\frac{\vec \sigma}{c}\right)\!\! =
	g_y(\vec p) + \frac{1}{c}
	\!\sum\limits_{\alpha}\! \Pder{g_y(\vec p)}{p_{\alpha}}(\sigma_{\alpha} - c p_{\alpha})
	+ \frac{1}{2c^2}
	\!\sum\limits_{\alpha\beta}\! (\sigma_{\alpha} - c p_{\alpha})(\sigma_{\beta} - c p_{\beta})
	\Psecder{g_y(\vec p)}{p_{\alpha}}{p_{\beta}} + \ldots
	\label{eq:taylorofg}
\end{equation}
We introduce the notation $K$ for the (scaled) covariance matrix of 
the multinomial-distributed $\vec \Sigma$,
\begin{align}
	K_{\alpha\beta} &\triangleq 
	\frac{1}{c} \cov{\Sigma_{\alpha}}{\Sigma_{\beta}} =  \delta_{\alpha\beta}p_{\alpha} - p_{\alpha}p_{\beta}.\label{eq:definitionofK}
\end{align}
For $\tau_{y \mid \vec p}$ we then get
\begin{equation}
	\tau_{y \mid \vec p} = \expec{\vec\Sigma|\vec p}{g_y\left(\frac{\vec \Sigma}{c}\right)}
	= g_y(\vec p) + \frac{1}{2c} \sum\limits_{\alpha\beta}
	K_{\alpha\beta} \Psecder{g_y(\vec p)}{p_{\alpha}}{p_{\beta}} +
	{\cal O}\left(\frac{1}{c\sqrt c}\right).
\label{eq:tayloroftau}
\end{equation}
The term containing the 1st derivative disappears because
$\expec{\vec\Sigma|\vec p}{\vec \Sigma - c \vec p} = 0$.
The ${\cal O}(1/c\sqrt c)$ comes from the fact that
$\left(\vec \Sigma - c \vec p\right)^n$ with $n \geq 2$
yields a result of order $c^{n/2}$ when the expectation over $\vec \Sigma$ is taken.
Now we have all the ingredients to do an expansion of $I(Y; \vec \Sigma \mid \vec P = \vec p)$
in terms of powers of $\frac{1}{c}$.
The details are given in Appendix~\ref{app:taylorexpansionofI}.
\begin{align}
	I(Y; \vec \Sigma \mid \vec P = \vec p) &= \frac{T(\vec p)}{2c\ln q}
	+  {\cal O}\left(\frac{1}{c \sqrt c}\right)
	\\
	T(\vec p) & \triangleq \sum\limits_y \frac{1}{g_y(\vec p)} \sum\limits_{\alpha\beta}
	K_{\alpha\beta} \Pder{g_y(\vec p)}{p_{\alpha}}\Pder{g_y(\vec p)}{p_{\beta}}.
\label{eq:defS}
\end{align}
Note that $T(\vec p$) can be related to Fisher Information.\footnote{
We can write
$T(\vec p)=\Tr [K(\vec p)\mathcal{I}(\vec p)]$, with $\mathcal{I}$ the Fisher information
of $Y$ conditioned on the $\vec p$ vector,
$\mathcal{I}_{\alpha\beta}(\vec p)\triangleq
\sum\limits_y g_y(\vec p) \left(\Pder{\ln g_y(\vec p)}{p_{\alpha}}\right)
\left(\Pder{\ln g_y(\vec p)}{p_{\beta}}\right)$.
}
The asymptotic fingerprinting game for $c \to \infty$ can now be stated as
\begin{align}
C_c(q) &= \frac{1}{2c^2 \ln q} \max_F \min_{\vec g} \int F(\vec p) T(\vec p) d^q \vec p.
\end{align}

\subsection{Change of variables}

Substitution of $K$ (\ref{eq:definitionofK}) into (\ref{eq:defS}) gives
\begin{equation}
	T(\vec p)=
	\sum\limits_y \frac{1}{g_y(\vec p)} \left\{ \sum\limits_{\alpha} p_{\alpha}
	\left( \Pder{g_y(\vec p)}{p_{\alpha}} \right)^2 - \left(\sum\limits_{\alpha} p_{\alpha}
	\Pder{g_y(\vec p)}{p_{\alpha}}\right)^2 \right\}.
\end{equation}

Now we make a change of variables
$p_\alpha=u_\alpha^2$ and $g_\alpha(\vec p)=\gamma_\alpha^2(\vec u)$,
with $u_\alpha\in[0,1]$, $\gamma_\alpha(\vec u)\in[0,1]$.
The hyperplane $\sum_\alpha p_\alpha=1$ becomes the hypersphere
$\sum_\alpha u_\alpha^2=1$.
For $\vec u$ on the hypersphere we must have $\sum_\alpha \gamma_\alpha^2(\vec u)=1$.
Due to the Marking Assumption, $u_\alpha=0$ implies $\gamma_\alpha(\vec u)=0$.
The change of variables induces the probability distribution $\Phi(\vec u)$
on the variable $\vec u$,
\begin{align}
	\Phi(\vec u) &\triangleq F(\vec p(\vec u)) \prod_{\alpha}(2 u_{\alpha}).
\end{align}
In terms of the new variables we have a much simplified expression,
\begin{align}
	T(\vec u) &= \sum\limits_y \left\{ \|\nabla\gamma_y\|^2
	- \left(\vec u \cdot \nabla \gamma_y \right)^2 \right\}.
\label{eq:Jasfunctionofu}
\end{align}
where $\nabla$ stands for the gradient $\partial/\partial\vec u$.

\subsection{Choosing $\vec \gamma$ outside the hypersphere}
\label{sec:freedominchoosinggamma}

The function $\vec g(\vec p)$ was introduced on the hypersphere $\sum_\alpha p_\alpha=1$,
but taking derivatives $\partial/\partial p_\alpha$ forces us to define $\vec g$ elsewhere
too.
In the new variables this means we have to define $\vec\gamma(\vec u)$
not only on the hypersphere `surface' $\|\vec u\|=1$ but also just outside of this surface.
Any choice will do, as long as it is sufficiently smooth.
A very useful choice is to make $\vec\gamma$ independent of $\|\vec u\|$,
i.e. dependent only on the `angular' coordinates in the surface. Then
we have the nice property $\vec u \cdot \nabla \gamma_y = 0$ for all $y \in \mathcal{Q}$,
so that (\ref{eq:Jasfunctionofu}) simplifies to
\begin{align}
	T(\vec u) &=  \sum_\alpha \| \nabla \gamma_\alpha \|^2 \label{eq:simplerJasfunctionofu}
\end{align}
and the asymptotic fingerprinting game to
\begin{align}
	C_c(q) &= \frac{1}{2c^2 \ln q} \max_{\Phi} \min_{\vec \gamma}
	\int \Phi(\vec u) T(\vec u) d^q \vec u.
\label{eq:gameintermsofu}
\end{align}

\subsection{Huang and Moulin's next step}
At this point \cite{HuangMoulin}
proceeds by applying the Cauchy-Schwartz inequality in a very clever way.
In our notation this gives
\begin{align}
	\max_{\Phi} \min_{\vec \gamma} \int \Phi(\vec u) T(\vec u) d^q \vec u &
	\geq 
	\max_{\Phi} \frac{1}{\int \frac{1}{\Phi(\vec u)} d^q \vec u} 
	\min_{\vec \gamma}[\int \sqrt{T(\vec u)} d^q \vec u]^2,
\label{CauchySchwartz}
\end{align}
with equality when $T$ is proportional to $1/\Phi^2$.
For the binary alphabet ($q=2$),
the integral $\int \sqrt{T(\vec u)} d^q \vec u$ becomes a known constant independent 
of the strategy~$\vec \gamma$. 
That makes the minimization over $\vec \gamma$ to disappear: 
The equality in (\ref{CauchySchwartz}) can then be achieved and
the entire game can be solved, yielding the arcsine bias distribution and interleaving attack
as the optimum.
For $q\geq 3$, however, the integral becomes dependent on the strategy $\vec \gamma$,
and the steps of \cite{HuangMoulin} cannot be applied.

\section{Asymptotic solution of the alternative game}
\label{sec:solution}

Our aim is to solve the alternative game to~(\ref{eq:gameintermsofu}), see Section~\ref{sec:alternative}.
\begin{align}
	C_c(q) &= \frac{1}{2c^2 \ln q} \min\limits_{\vec \gamma} \max\limits_{\vec u} T(\vec u)
\label{eq:alternativegameintermsofu}.
\end{align}
First we prove a lower bound on $\max_{\vec u} T(\vec u)$ for any strategy $\vec \gamma$.
Then we show the existence of a strategy which attains this lower bound.
The first part of the proof is stated in the following theorem.

\begin{theorem}
\label{the:lowerboundonalternativegame}
For any strategy $\vec \gamma$ satisfying
the Marking Assumption ($u_\alpha\!=\!0\implies\gamma_\alpha(\vec u)\!=\!0$)
and conservation of probability
($\|\vec u\|=1 \implies \|\vec\gamma(\vec u)\|=1$)
the following inequality holds:
\begin{align}
	\max\limits_{\vec u:\; \vec u\geq 0, \|\vec u\|=1} 
	T(\vec u) & \geq q - 1.
\end{align}
\end{theorem}

\underline{\it Proof:}
We start with the definition of the Jacobian matrix $J(\vec u)$:
\begin{align}
	J_{\alpha\beta}(\vec u) & \triangleq \Pder{\gamma_{\alpha}(\vec u)}{u_{\beta}}.
\end{align}
In this way we can write:
\begin{align}
	T(\vec u) &= \Tr(J^T J).
\end{align}
The matrix $J$ has rank at most $q -1$, because of our
choice
$\vec u \cdot \nabla \gamma_y = 0$
which can be rewritten as $J \vec u = 0$.
That implies that the rank of $J^T J$ is also at most $q - 1$.
Let $\lambda_1(\vec u), \lambda_2(\vec u), \ldots,\lambda_{q-1}(\vec u)$ be the nonzero eigenvalues of $J^T J$.
Then
\begin{align}
	T(\vec u) &= \sum\limits_{i = 1}^{q - 1} \lambda_i(\vec u).
\end{align}
Let $\vec v_1, \vec v_2, \ldots, \vec v_{q-1}$ be the unit-length eigenvectors of $J^T J$ and let
$\rd\vec u_{(1)}$, $\rd \vec u_{(2)}$, $\ldots$, $\rd \vec u_{(q-1)}$ be infinitesimal displacements in the directions of these eigenvectors, i.e. $\rd \vec u_{(i)}\propto \vec v_i$.
According to Lemma~\ref{lem:symmetricmatrix} the eigenvectors are mutually orthogonal. Thus we can write
the $(q-1)$-dimensional `surface' element $\rd S_{\vec u}$ of the hypersphere in terms of these displacements:
\begin{align}
	\rd S_{\vec u} &= \prod\limits_{i = 1}^{q-1} \|\rd \vec u_{(i)}\|.
\end{align}
Any change $\rd\vec u$ results in a change $\rd\vec\gamma=J\rd\vec u$.
Hence we have
$\rd \vec \gamma_{(i)} = J \rd \vec u_{(i)}$.
By Lemma~\ref{lem:symmetricmatrix}, the displacements $\rd \vec \gamma_{(1)}, \rd \vec \gamma_{(2)}, \ldots, \rd \vec \gamma_{(q-1)}$ are mutually orthogonal and we can express the 
$(q-1)$-dimensional 
`surface' element $\rd S_{\vec \gamma}$ as
\begin{align}
\rd S_{\vec \gamma} &= \prod\limits_{i=1}^{q-1} \|\rd \vec \gamma_{(i)} \| = \prod\limits_{i=1}^{q-1} \sqrt{\|J \rd \vec u_{(i)} \|^2}\\
&= \prod\limits_{i=1}^{q-1} \sqrt{\rd \vec u_{(i)}^T J^T J \rd \vec u_{(i)}} = \prod\limits_{i=1}^{q-1} \| \rd \vec u_{(i)} \| \sqrt{\lambda_i} \\
&= \rd S_{\vec u} \prod\limits_{i=1}^{q-1} \sqrt{\lambda_i}.
\end{align}
We define
the spatial average over $\vec u$ as
Av$_u[f(\vec u)]\triangleq \int\! f(\vec u)\; \rd S_u/ \int\! \rd S_u$.
We then have

\begin{align}
{\rm Av}_u[\sqrt{\lambda_1 \lambda_2 \dots \lambda_{q-1}}]  = \frac{\int dS_{\vec u} \sqrt{\lambda_1 \lambda_2 \dots \lambda_{q-1}}}{\int dS_{\vec u}}
= \frac{\int dS_{\vec \gamma}}{\int dS_{\vec u}} \geq 1
\end{align}
where the inequality follows from Lemma~\ref{lem:continuousmapping} applied to the mapping
$\vec \gamma(\vec u)$.
(The hypersphere orthant $\|\vec u\|=1$, $\vec u\geq 0$ is closed and contains no holes;
the $\vec\gamma$ was defined as being twice differentiable; 
the edge of the hypersphere orthant is given by the pieces where $u_i=0$ for some $i$;
these pieces are mapped to themselves due to the Marking Assumption.
The edges of the edges are obtained by setting further components of $\vec u$ to zero, etc.
Each of these sub-edges is also mapped to itself due to the Marking Assumption.
In the one-dimensional sub-sub-edge we apply the intermediate value theorem, which proves
surjectivity. From there we recursively apply Lemma~\ref{lem:continuousmapping}
to increasing dimensions, finally reaching dimension $q-1$).

Since the spatial average is greater than or equal to $1$ there must exist a point 
$\vec u_*$ where \\
$\sqrt{\lambda_1(\vec u_*) \lambda_2(\vec u_*) \dots\lambda_{q-1}(\vec u_*)} \geq 1$.
Now we apply Lemma~\ref{lem:amgminequality},
\begin{align}
	T(\vec u_*)=\sum\limits_{i=1}^{q-1} \lambda_i(\vec u_*) &\geq (q-1)
	\sqrt[q-1]{\lambda_1(\vec u_*)\lambda_2(\vec u_*)\dots\lambda_{q-1}(\vec u_*)} \geq q-1.
\end{align}
The last inequality holds since $\sqrt x\geq 1$ implies $\sqrt[q-1]x\geq 1$.
Finally $\max_{\vec u} T(\vec u) \geq T(\vec u_*) \geq q - 1$.
\hfill$\square$

Next we show the existence of a strategy which attains this lower bound.

\begin{theorem}
\label{the:interleavingachieveslowerbound}
Let the interleaving attack $\vec \gamma$ be extended beyond the hypersphere
$\|\vec u\|=1$ as
$\gamma_y(\vec u) = \frac{u_y}{\|u\|}$, satisfying $\vec u\cdot\nabla\gamma_y=0$ for all $y$.
For the interleaving attack we then have $T(\vec u) = q - 1$
for all $\vec u \geq 0, \| \vec u \| = 1$.
\end{theorem}
\underline{\it Proof:}
\begin{align}
\Pder{\gamma_y(\vec u)}{u_{\alpha}} &= \frac{\delta_{y\alpha}}{\|u\|} - \frac{u_y u_{\alpha}}{\|u\|^3}.\\
T(\vec u) &= \sum\limits_y \|\nabla \gamma_y(\vec u)\|^2 = \sum\limits_y\sum\limits_{\alpha}\left(\frac{\delta_{y\alpha}}{\|u\|} - \frac{u_y u_{\alpha}}{\|u\|^3}\right)^2\\
&= \sum\limits_y \left\{\frac{1}{\|u\|^2} - \frac{u_y^2}{\|u^4\|} \right\} = \frac{q-1}{\|u\|^2}
\end{align}
where we used the property $\delta_{y\alpha}^2 = \delta_{y\alpha}$.
For $\| u \| = 1$ it follows that $T(\vec u) = q - 1$.
\hfill$\square$

These two theorems together give the solution of the min-max game~(\ref{eq:alternativegameintermsofu}). 
The main result of this paper is stated in the following theorem:

\begin{theorem}
\label{the:mainresult}
The asymptotic fingerprinting capacity $C_c^\infty(q)$ 
in the limit $c \to \infty$ for an alphabet of size $q$ is given by
\begin{align}
	C_c^\infty(q) &= \frac{q - 1}{2c^2 \ln q}.
\end{align}
\end{theorem}

\underline{\it Proof:}
For any strategy $\vec \gamma$, Theorem~\ref{the:lowerboundonalternativegame} shows that
$\max_{\vec u} T(\vec u) \geq q - 1$. As shown in Theorem~\ref{the:interleavingachieveslowerbound},
the interleaving attack has $T(\vec u) = q - 1$ independent of $\vec u$. Hence
\begin{align}
	\min\limits_{\vec \gamma} \max\limits_{\vec u} T(\vec u) & = q  - 1
\end{align}
is the solution of the min-max game. By Sion's theorem this is also the pay-off solution to the
max-min game, as shown in Section~\ref{sec:alternative}.
Substitution into
(\ref{eq:alternativegameintermsofu}) yields the final result.
\hfill$\square$

{\it Remark:}
Any distribution function $\Phi(\vec u)$ is an optimum of the minimax game,
since $T(\vec u)$ is constant for the optimal choice of $\vec\gamma$.

\section{Discussion}
\label{sec:discussion}

We have proven that the asymptotic channel capacity is 
$C_c^\infty(q)=\frac{q-1}{c^2 2\ln q}$.
This is an increasing function of $q$; hence there is an advantage
in choosing a large alphabet whenever the details of the watermarking system allow it.

Some confusion may arise
because of the difference between binary and $q$-ary symbols,
and the `space' they occupy in content. 
Therefore we explicitly mention the following.
The capacity is an upper bound on the achievable rate of (reliable) codes, where the
rate measures which fraction of the occupied `space' confers actual information.
The higher the fraction, the better, independent of the nature of the symbols.
Thus the rate (and channel capacity) provides 
a fair comparison between codes that have different~$q$.

The obvious next question is how to construct a $q$-ary scheme that achieves
capacity.
We expect that a straightforward generalization of the Amiri-Tardos scheme
\cite{AmiriTardos} will do it.
Constructions with more practical accusation algorithms, like \cite{symmetric},
do not achieve capacity but have already shown that non-binary codes achieve
higher rates than their binary counterparts.

\vskip2mm

When it comes to increasing $q$, one has to be cautious
for various reasons.
\begin{itemize}
\item[$\bullet$]
The actually achievable value of $q$ is determined by 
the watermark embedding technique and
the attack mechanism at the signal processing level.
Consider for instance a $q=8$ code implemented in such a way that a $q$-ary symbol
is embedded in the form of three parts (bits) that can be attacked independently.
Then the Marking Assumption will no longer
hold in the $q=8$ context, and the
`real' alphabet size is in fact $2$. 
\item[$\bullet$]
A large $q$ can cause problems for accusation schemes that use
an accusation sum as in \cite{symmetric}.
As long as the probability distributions of the accusation sums
are approximately Gaussian, the accusation works well.
It was shown in \cite{TardosFourier}
that increasing $q$ causes the tails of the probability distribution
to slowly become less Gaussian, which is bad for the code rate.
On the other hand, the tails become more Gaussian with increasing $c$.
This leads us to believe that for this type of accusation there is an optimal $q$
as a function of~$c$.
\end{itemize}
The proof technique used in this paper does not reveal
the asymptotically optimal bias distribution and attack strategy.
This is left as a subject for future work.

\vskip2mm

{\bf Acknowledgements}\\
Discussions with Jan de Graaf, Antonino Simone, Jan-Jaap Oosterwijk and Benne de Weger are
gratefully acknowledged. We thank Teddy Furon for calling our attention to
the Fisher Information.


\bibliographystyle{plain}
\bibliography{capacity}

\begin{thebibliography}{10}

\bibitem{AmiriTardos}
E.~Amiri and G.~Tardos.
\newblock High rate fingerprinting codes and the fingerprinting capacity.
\newblock In {\em ACM-SIAM Symposium on Discrete Algorithms (SODA) 2009}, pages
  336--345.

\bibitem{ABD}
N.P. Anthapadmanabhan, A.~Barg, and I.~Dumer.
\newblock Fingerprinting capacity under the marking assumption.
\newblock {\em IEEE Transaction on Information Theory -- Special Issue on
  Information-theoretic Security}, 54(6):2678--2689.

\bibitem{BlayerTassa}
O.~Blayer and T.~Tassa.
\newblock Improved versions of {Tardos}' fingerprinting scheme.
\newblock {\em Designs, Codes and Cryptography}, 48(1):79--103, 2008.

\bibitem{FuronEM}
A.~Charpentier, F.~Xie, C.~Fontaine, and T.~Furon.
\newblock Expectation maximization decoding of {Tardos} probabilistic
  fingerprinting code.
\newblock In {\em Media Forensics and Security 2009}, page 72540.

\bibitem{CoverThomas}
T.M. Cover and J.A. Thomas.
\newblock {\em Elements of information theory}.
\newblock Wiley Series in Telecommunications. Wiley \& Sons, 1991.

\bibitem{HuangMoulin}
Y.-W. Huang and P.~Moulin.
\newblock Maximin optimality of the arcsine fingerprinting distribution and the
  interleaving attack for large coalitions.
\newblock In {\em IEEE Workshop on Information Forensics and Security (WIFS)
  2010}.

\bibitem{HM_ISIT2009}
Y.-W. Huang and P.~Moulin.
\newblock Saddle-point solution of the fingerprinting capacity game under the
  marking assumption.
\newblock In {\em {IEEE International Symposium on Information Theory (ISIT)
  2009}}, pages 2256--2260.

\bibitem{Moulin2008}
P.~Moulin.
\newblock Universal fingerprinting: Capacity and random-coding exponents.
\newblock In {\em {IEEE International Symposium on Information Theory (ISIT)
  2008}}, pages 220--224.
\newblock \url{http://arxiv.org/abs/0801.3837v2}.

\bibitem{Nuida_DCC2009}
K.~Nuida, S.~Fujitsu, M.~Hagiwara, T.~Kitagawa, H.~Watanabe, K.~Ogawa, and
  H.~Imai.
\newblock An improvement of discrete {Tardos} fingerprinting codes.
\newblock {\em Designs, Codes and Cryptography}, 52(3):339--362, 2009.

\bibitem{Nuida}
K.~Nuida, M.~Hagiwara, H.~Watanabe, and H.~Imai.
\newblock Optimal probabilistic fingerprinting codes using optimal finite
  random variables related to numerical quadrature.
\newblock {\em CoRR}, abs/cs/0610036, 2006.

\bibitem{TardosFourier}
A.~Simone and B.~\v{S}kori\'{c}.
\newblock Accusation probabilities in {Tardos} codes.
\newblock {\it Benelux Workshop on Information and System Security (WISSEC)
  2010.} \url{http://eprint.iacr.org/2010/472}.

\bibitem{Sion}
M.~Sion.
\newblock On general minimax theorems.
\newblock {\em Pacific Journal of Mathematics}, 8(1):171--176, 1958.

\bibitem{Tardos}
G.~Tardos.
\newblock Optimal probabilistic fingerprint codes.
\newblock In {\em {STOC} 2003}, pages 116--125.

\bibitem{symmetric}
B.~\v{S}kori\'{c}, S.~Katzenbeisser, and M.U. Celik.
\newblock Symmetric {Tardos} fingerprinting codes for arbitrary alphabet sizes.
\newblock {\em Designs, Codes and Cryptography}, 46(2):137--166, 2008.

\bibitem{CombinedDigit}
B.~\v{S}kori\'{c}, S.~Katzenbeisser, H.G. Schaathun, and M.U. Celik.
\newblock {Tardos fingerprinting codes in the combined digit model}.
\newblock In {\em IEEE Workshop on Information Forensics and Security (WIFS)
  2009}, pages 41--45.

\bibitem{SVCT}
B.~\v{S}kori\'{c}, T.U. Vladimirova, M.U. Celik, and J.C. Talstra.
\newblock Tardos fingerprinting is better than we thought.
\newblock {\em IEEE Trans. on Inf. Theory}, 54(8):3663--3676, 2008.

\bibitem{XFF}
F.~Xie, T.~Furon, and C.~Fontaine.
\newblock On-off keying modulation and {Tardos} fingerprinting.
\newblock In {\em {MM{\&}Sec} 2008}, pages 101--106.

\end{thebibliography}

\appendix

\section{Taylor expansion of $I(Y; \vec \Sigma \mid \vec P = \vec p)$}
\label{app:taylorexpansionofI}
We compute the leading order term of $I(Y; \vec \Sigma \mid \vec P = \vec p)$ from~(\ref{eq:mutualinformationintermsofg}) with respect to powers of $\frac{1}{c}$. 
We write $\log_q g_y = \ln g_y/\ln q$ and, using (\ref{eq:taylorofg}),
$\ln g_y(\vec\sigma/c)=\ln[g_y(\vec p)+\epsilon_y]=\ln g_y(\vec p)+\ln(1+\epsilon_y/g_y(\vec p))$,
where we have introduced the shorthand notation
\begin{equation}
	\epsilon_y \triangleq \frac{1}{c}\sum\limits_{\alpha} 
	\Pder{g_y(\vec p)}{p_{\alpha}}(\Sigma_{\alpha} - c p_{\alpha})
	+ \frac{1}{2c^2} \sum\limits_{\alpha\beta} (\Sigma_{\alpha} 
	- c p_{\alpha})(\Sigma_{\beta} - c p_{\beta})\Psecder{g_y(\vec p)}{p_{\alpha}}{p_{\beta}} + \ldots
\label{eq:defepsilon}
\end{equation}
Higher derivative terms are omitted since they contain higher powers of $1/c$ (even after the expectation
over $\vec\Sigma$ is taken).
Next we apply the Taylor expansion
$\ln (1 + x) = x -\frac{x^2}{2} + \cdots$, resulting in
\begin{align}
	\ln g_y (\frac{\vec \Sigma}{c}) 
&= \ln g_y(\vec p) + \frac{\epsilon_y}{g_y(\vec p)} - \frac{\epsilon_y^2}{2g_y^2(\vec p)} + \ldots
\end{align}

where we stop after the second order term since that is already of order $\frac{1}{c}$ 
when we take the expectation over $\vec \Sigma$. 
Using~(\ref{eq:tayloroftau}) we get
\begin{align}
\ln \tau_{y \mid \vec p} &= \ln g_y(\vec p) + \frac{\zeta_y}{g_y(\vec p)} + \ldots,\\
\zeta_y &\triangleq \frac{1}{2c} \sum\limits_{\alpha\beta} K_{\alpha\beta} 
\Psecder{g_y(\vec p)}{p_{\alpha}}{p_{\beta}} + {\cal O}\left(\frac{1}{c\sqrt{c}}\right)
\end{align}
Now we combine all the ingredients,
\begin{align}
g_y \left(\frac{\vec \Sigma}{c}\right) 
\ln \left(\frac{g_y\left(\frac{\vec \Sigma}{c}\right)}{\tau_{y \mid \vec p}}\right) 
&= \left(g_y(\vec p) + \epsilon_y + \ldots \right)\left(\frac{\epsilon_y - \zeta_y}{g_y(\vec p)} 
- \frac{\epsilon_y^2}{2g_y^2(\vec p)} + \ldots \right) 
\label{eq:expansionofglogg}
\end{align}
where in the first factor we stop at $\epsilon_y$ 
because when the expectation over $\vec \Sigma$ is applied, $\epsilon_y^2$ 
gives at least a factor of $\frac{1}{c}$ 
and the terms in the second factor give at least a factor of $\frac{1}{\sqrt c}$.

Now $\expec{\vec \Sigma \mid \vec P = \vec p}{\epsilon_y - \zeta_y} = 0$ 
because $\expec{\vec \Sigma \mid \vec P = \vec p}{\vec \Sigma - c\vec p} = 0$ 
and $\zeta_y$ was defined as the expectation over $\vec \Sigma$ 
of the second term in~(\ref{eq:defepsilon}). 
The expectation of the product $\expec{\vec \Sigma \mid \vec P = \vec p}{\epsilon_y \zeta_y}$ 
is of order $\frac{1}{c^2}$ and so we drop it as well.  
The only remaining part of order $\frac{1}{c}$ in~(\ref{eq:expansionofglogg}) 
is $\frac{\epsilon_y^2}{2 g_y(\vec p)}$ and hence we end up with:

\begin{align}
& I(Y; \vec \Sigma \mid \vec P = \vec p) \nonumber\\
&= \frac{1}{2\ln q} \sum_y \frac{1}{g_y(\vec p)} \expec{\vec \Sigma \mid \vec P = \vec p}{\epsilon_y^2} + 
{\cal O}\left(\frac{1}{c\sqrt c}\right) \\
&= \frac{1}{2c^2 \ln q} \sum_y \frac{1}{g_y(\vec p)}\expec{\vec \Sigma \mid \vec P = \vec p}
{\left(\sum_{\alpha} \Pder{g_y(\vec p)}{p_{\alpha}}(\Sigma_{\alpha} - c p_{\alpha})\right)^2} 
+ {\cal O}\left(\frac{1}{c\sqrt c}\right) \\
&= \frac{1}{2c \ln q} \sum_y \frac{1}{g_y(\vec p)} 
\sum_{\alpha\beta} K_{\alpha\beta} \Pder{g_y(\vec p)}{p_{\alpha}}\Pder{g_y(\vec p)}{p_{\beta}} 
+ {\cal O}\left(\frac{1}{c\sqrt c}\right)
\end{align}

where in the second step we expanded $\epsilon_y^2$ and took the square of only the first term in~(\ref{eq:defepsilon}) because the other combination of terms give rise to higher powers of $\frac{1}{c}$.


\end{document}